\documentclass[twoside,british,a4paper,twocolumn]{article}

\usepackage{amsmath}
\usepackage{amsfonts}
\usepackage{amssymb,upref}
\usepackage{tgheros} 
\usepackage{tgtermes} 
\usepackage{anyfontsize}
\usepackage{xcolor}
\usepackage{graphicx}
\usepackage{enumitem}
\usepackage{subfig}
\usepackage{siunitx}
\usepackage{tikz}

\usepackage{balance}

\usepackage[utf8]{inputenc}
\usepackage[T1]{fontenc}

\usepackage{isodate}
\cleanlookdateon

\usepackage{lettrine}
\pdfmapfile{=montserrat.map} 

\usepackage{lipsum}

\usepackage{newtxmath}

\usepackage{geometry}
\usepackage{fancyhdr}

\fancypagestyle{firstpagestyle}
\usepackage{titlesec}
\titleformat{\section}[block]{\bfseries\upshape\sffamily\boldmath}{}{0.em}{}
\titlespacing*{\section}{0pt}{0.8em plus 0ex minus 0ex}{0em plus 0.ex}

\usepackage{titling}
\pretitle{\noindent\begin{minipage}{\textwidth} \raggedright\huge\bfseries\upshape\sffamily\boldmath
\color[rgb]{0,0,0.8}
}
\posttitle{\end{minipage} \vskip 0.6em}
\preauthor{\noindent \begin{minipage}{\textwidth} \large\mdseries\sffamily}
\postauthor{
  \end{minipage} 
  \vskip 0.4em
  {
   \sffamily
   \color{black!80}
   \noindent \small
   \address
   \par 
   \authoremail
   }
   \par \vskip 0.4em
  }
\predate{ \noindent \hspace{-0.4em} \sffamily\small}
\postdate{\vskip 0.0em}

\usepackage{caption}
\DeclareCaptionFont{cfs}{\fontsize{8.5}{10.25}\selectfont}
\DeclareCaptionLabelSeparator{vline}{\;|\;}
\usepackage{tcolorbox}
\definecolor{abstractboxcolor}{cmyk}{0.1,0,0,0}
\newtcolorbox{abstractbox}{
  arc=0pt,
  boxrule=0pt,
  colback=abstractboxcolor,
  boxsep=0.5em,
  left=0pt, right=0pt, bottom=0pt, top=0pt,
  width=\columnwidth
}
\makeatletter
 \def\@textbottom{\vskip \z@ \@plus 1pt}
 \let\@texttop\relax
\makeatother
\usepackage[super,numbers,sort&compress]{natbib}

\makeatletter \def\NAT@def@citea{\def\@citea{\NAT@separator\,}} \makeatother 
\usepackage{etoolbox}
\apptocmd{\sloppy}{\hbadness 10000\relax}{}{}

\definecolor{orcidgreen}{rgb}{0.65098041,0.80784315,0.22352941}

\newcommand{\orcidicon}[1]{%
\resizebox{#1}{!}{%
\begin{tikzpicture}[scale=0.010]%
\fill [orcidgreen]
  (256,128)
    .. controls (256,57.3)  and (198.7,0)   .. (128,0)
    .. controls (57.3,0)    and (0,57.3)    .. (0,128) 
    .. controls (0,198.7)   and (57.3,256)  .. (128,256)
    .. controls (198.7,256) and (256,198.7) .. (256,128)
    ;

\fill [white]
  (86.3,69.8)--(70.9,69.8)
             --(70.9,176.9)
             --(86.3,176.9)
             --(86.3,128.5)
             --(86.3,69.8)
  ;

\fill [white]
  (88.7,199.2)
    .. controls (88.7,193.7) and (84.2,189.1) .. (78.6,189.1)
    .. controls (73,189.1)   and (68.5,193.7) .. (68.5,199.2)
    .. controls (68.5,204.8) and (73,209.3)   .. (78.6,209.3)
    .. controls (84.2,209.3) and (88.7,204.7) .. (88.7,199.2)
  ;

\fill [white]
  (108.9,176.9)
  --(150.5,176.9)
  .. controls (190.1,176.9) and (207.5,148.6) ..(207.5,123.3)
  .. controls (207.5,95.8) and (186,69.7) .. (150.7,69.7)
  --(108.9,69.7)
  --(108.9,176.9)
  (124.3,83.6)
  --(148.8,83.6)
  .. controls (183.7,83.6)  and (191.7,110.1) .. (191.7,123.3)
  .. controls (191.7,144.8) and (178,163)     .. (148,163)
  --(124.3,163)
  --(124.3,83.6)
  ;
\end{tikzpicture}%
}%
}

\newlength{\orcidiconwidth}
\newcommand{\authororcid}[1]{%
  \hspace{0.5\orcidiconwidth}%
  \href{https://orcid.org/#1}{\orcidicon{\orcidiconwidth}}%
  }

\usepackage[%
  bookmarks=true,
  colorlinks,
  linkcolor=blue,
  urlcolor=blue,
  citecolor=blue,
  plainpages=false,
  pdfpagelabels,
  final,
  breaklinks=true
]{hyperref}
\hypersetup{
pdftitle={Generation of optical Schrödinger cat states in intense laser-matter interactions}, 
pdfauthor={M. Lewenstein, M.F. Ciappina, E. Pisanty, J. Rivera-Dean, P. Stammer, Th. Lamprou, and P. Tzallas}
}

\title{Generation of optical Schr\"{o}dinger cat states in intense laser-matter interactions}
\newcommand\shorttitle{Generation of optical Schr\"{o}dinger cat states in intense laser-matter interactions}

\author{\raggedright
  M.~Lewenstein\authororcid{0000-0002-0210-7800}$^\mathsf{1,2\ast}$, 
  M.F.~Ciappina\authororcid{0000-0002-1123-6460}$^\mathsf{1,3,4,5}$, 
  E.~Pisanty\authororcid{0000-0003-0598-8524}$^\mathsf{1,6}$, 
  J.~Rivera-Dean\authororcid{0000-0003-3031-0029}$^\mathsf{1}$, 
  P.~Stammer$^\mathsf{1,6}$, 
  Th.~Lamprou$^\mathsf{7,8}$, and 
  P.~Tzallas\authororcid{0000-0002-8063-5596}$^\mathsf{7,9\ast}$
  }
\newcommand\shortauthor{M. Lewenstein et al.}

\newcommand\address{
  \textsuperscript{1}ICFO -- Institut de Ciencies Fotoniques, The Barcelona Institute of Science and Technology, 08860 Castelldefels (Barcelona), Spain\\
  \textsuperscript{2}ICREA, Pg. Llu\'{\i}s Companys 23, 08010 Barcelona, Spain\\
  \textsuperscript{3}Institute of Physics of the ASCR, ELI-Beamlines project, Na Slovance 2,182 21 Prague, Czech Republic\\
  \textsuperscript{4}Physics Program, Guangdong Technion -- Israel Institute of Technology, Shantou, Guangdong 515063, China\\
  \textsuperscript{5}Technion -- Israel Institute of Technology, Haifa, 32000, Israel\\
  \textsuperscript{6}Max Born Institute for Nonlinear Optics and Short Pulse Spectroscopy, Max Born Strasse 2a, D-12489 Berlin, Germany\\
  \textsuperscript{7}Foundation for Research and Technology-Hellas, Institute of Electronic Structure \& Laser, GR-70013 Heraklion (Crete), Greece\\
  \textsuperscript{8}Department of Physics, University of Crete, P.O. Box 2208, GR-71003 Heraklion (Crete), Greece\\
  \textsuperscript{9}ELI-ALPS, ELI-Hu Non-Profit Ltd., Dugonics tér 13, H-6720 Szeged, Hungary
  }
\newcommand\authoremail{$^\ast$Corresponding authors e-mail:  maciej.lewenstein@icfo.eu and ptzallas@iesl.forth.gr}

\date{ 19 August 2021 }

\usepackage{physics}

\usepackage{textcomp}

\begin{document}

\twocolumn[
\begin{@twocolumnfalse}

\maketitle
\thispagestyle{firstpagestyle}

\vspace{-2mm}

Accepted Manuscript for
\href{%
  https://doi.org/10.1038/s41567-021-01317-w
  }{%
  \textit{Nat.\ Phys.} \textbf{17}, 1104–1108 (2021)%
  },
available as
\href{%
  https://arxiv.org/abs/2008.10221
  }{%
  arXiv:2008.10221%
  }%
.

\vspace{2mm}

\end{@twocolumnfalse}
]

\lettrine[lines=3, lhang=0.15]{T}{\:} {\bf he
physics of intense laser--matter interactions\cite{Mourou, Strickland} is described by treating the light pulses classically, anticipating no need to access optical measurements beyond the classical limit. However, the quantum nature of the electromagnetic fields is always present\cite{Glauber2}. Here, we demonstrate that intense laser--atom interactions may lead to the generation of highly non-classical light states. This was achieved by using the process of high-harmonic generation in atoms\cite{McPherson1987,Ferray1988}, in which the photons of a driving laser pulse of infrared frequency are up-converted into photons of higher frequencies in the extreme ultraviolet spectral range. The quantum state of the fundamental mode after the interaction, when conditioned on the high-harmonic generation, is a so-called Schr\"{o}dinger cat state, which corresponds to a superposition of two distinct coherent states: the initial state of the laser and the coherent state reduced in amplitude that results from the interaction with atoms. The results open the path for investigations towards the control of the non-classical states, exploiting conditioning approaches on physical processes relevant to high-harmonic generation.
}

Decades after the invention of laser\cite{Maiman}, the development of high-power lasers\cite{Mourou, Strickland} enabled scientists to explore laser-matter interactions driven by laser--fields with strengths comparable or stronger than the atomic potential. This research domain, termed strong-laser-field physics, opened the way for studies ranging from relativistic optics\cite{Mourou}, to high--harmonic generation (HHG), and ultrafast optoelectronics \cite{Symphony}. A large amount of work was conducted in the past for the description of intense laser-atom interaction. In the early days, it was described perturbatively in quantized manner using multiphoton processes \cite{Delone, Protopapas} and, later on, in the strong-field limit by semi-classical approaches like the three-step model\cite{Corkum1993, Kulander1993,Lewenstein1994}. However, due to the high photon number of the driving--field, the description of these interactions in the strong-field limit remains incomplete as it relies on semi-classical approximations treating the atomic system quantum mechanically but the electromagnetic field classically i.e. ignoring the quantum nature of the electromagnetic radiation, which is always present. Revealing the quantum nature of light in strongly laser--driven interactions, besides its fundamental physical interest, is important for applications in basic research and technology. This is because it bridges the gap between strong laser--field physics\cite{Mourou, Strickland} and quantum optics\cite{Glauber2}, providing access to the development of strong-field quantum electrodynamics, and of a new class of non-classical light sources (like squeezed, Fock, Schr\"{o}dinger ``cat'' states, etc.)\cite{Knight, Ourjoumtsev, Zavata, Ourjoumtsev1, Wakui} which are at the core of quantum technology\cite{Acin, Walmsay, Deutsch}.

\begin{figure*}	
\begin{center}
\includegraphics[width=15 cm]{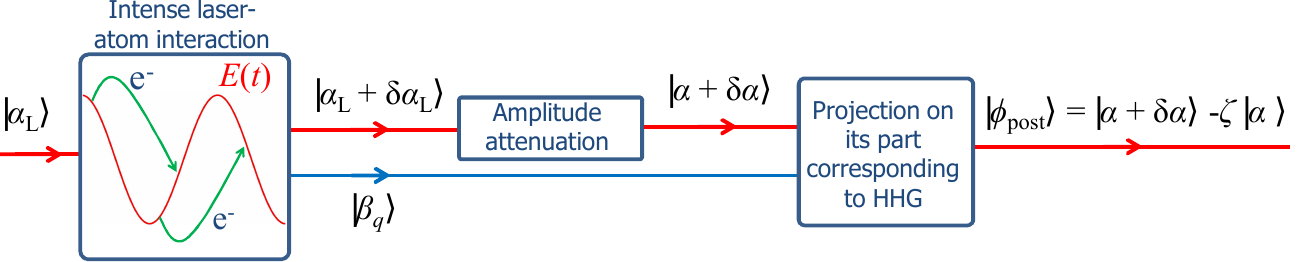}
\end{center}
\caption{\textbf{ Schematic representation of the generation of optical ``cat'' states.} The coherent laser state $|\alpha_L \rangle$ (thick red lines) interacts with atoms and in consequence high harmonics (thick blue lines) are generated. The inset shows an intuitive picture of the electron recollision process which leads to HHG. The oscillating laser field $E(t)$ is depicted with thin red curve and the electron paths with thin green curves. After the interaction the harmonics modes are coherent $|\beta_q\rangle$ and the fundamental is an amplitude--shifted coherent state $|\alpha_L +\delta\alpha_L\rangle$. This state after an amplitude attenuation ($|\alpha +\delta\alpha\rangle$), is projected on its part corresponding to HHG and becomes a Schr\"{o}dinger ``cat'' state $|\phi_{\rm post}\rangle$ with $\zeta=\langle\alpha|\alpha +\delta\alpha\rangle$.}
\end{figure*}

To unmask the quantum nature of light in strongly laser--driven interactions and show its impact on the aforementioned directions, we have used the HHG in atoms\cite{McPherson1987,Ferray1988}, which is one of the most fundamental processes in strong laser-field physics. The understanding of the HHG process was boosted from  the formulation of the semi-classical three-step model\cite{Lewenstein1994}. In this approach, HHG from a single atom/molecule/solid is initiated by electron's tunneling to the continuum, its subsequent acceleration in the intense laser field, and finally its recombination to the ground state of the target. To provide a fully quantized description of the HHG process, we have to rigorously answer the following questions: I) what is the quantum depletion of the coherent state of the fundamental laser mode? and II) what is the quantum state of the generated harmonics? Although several groups attempted to study this problem theoretically \cite{Diestler2008, Paris1, Foldi, Yangaliev, Gorlach2019} and experimentally \cite{Paris2,Paris3}, none of these attempts provides a rigorous answer to the above questions. Here, we show that if the initial state of a system of $N$ atoms in their ground state is impinged by a laser in a coherent state of amplitude $\alpha_L$,  then the resulting quantum states of the fundamental and harmonic modes are coherent. However, due to coherences and correlations between the fundamental and the harmonics, the fundamental mode amplitude is shifted (i.e. $\alpha_L \to \alpha_L + \delta \alpha_L$), where $\delta \alpha_L$ is negative and reflects the energy conservation which, in the context of the semi-classical three-step model \cite{Lewenstein1994}, represents the energy losses of the fundamental due to the re-collision process. While at the single atom level $\delta \alpha_L$ is negligibly small, it may become quite significant due to the cooperative effect in  HHG and the phase matched contribution of many atoms. Hence, the final state of the fundamental mode, conditioned on harmonic generation, is a superposition of the two coherent states: the initial and the shifted one. This is, in general, highly non-classical: it interpolates between a Schr\"{o}dinger ``kitten'' state, corresponding to a coherently shifted first Fock state (for  $|\langle \alpha_L + \delta \alpha_L| \alpha_L \rangle| \approx 1 $),  and a ``cat'' state (for $0<|\langle \alpha_L + \delta \alpha_L| \alpha_L \rangle| <1 $). We exactly calculate $\delta \alpha_L $, as well as the coherent amplitudes of the harmonics, and we confirm experimentally the generation of the ``cat'' state, by characterizing the quantum state of the fundamental field exiting the HHG medium when conditioned to HHG. This was achieved by combining the Quantum Tomography (QT) approach \cite{Knight, Lvovsky} with a photon--correlation--based method, namely the Quantum Spectrometer (QS) \cite{Paris2,Paris3}.

Starting from the time-dependent Schr\"odinger equation (TDSE), describing both the atom and light quantum mechanically, it can be shown (Methods) that the evolution of the state of the laser and harmonic modes ({\it q}), conditioned on the atomic ground state and upon neglecting of the (small) continuum/excited part, is described by
	\begin{equation}
		i\hbar\frac{\partial}{ \partial t} |\phi(t) \rangle=
		-\hat{\bf E}_Q(t)\cdot \langle\hat{\bf d}_H(t)\rangle|\phi(t) \rangle.
		\label{eqphi}
	\end{equation}
Here, $\hat{\bf E}_Q(t)$ describes the quantum fluctuating part of the laser electric-field and $\langle\hat{\bf d}_H(t)\rangle$ is the quantum averaged time-dependent dipole moment induced by the classical part of the laser pulse, which can be efficiently calculated solving the TDSE, or even easier using the strong field approximation (SFA) \cite{Lewenstein1994, Symphony}. The solution is (Methods):
	\begin{equation}
		|\phi(t) \rangle=|(\alpha_L + \delta\alpha_L) e^{-i\omega_Lt}, \beta_3e^{-3i\omega_Lt},\ldots,\beta_qe^{-qi\omega_Lt}, \ldots\rangle,
	\end{equation}
where $\delta\alpha_L=-i{\bf g}(\omega_L)\cdot {\bf d}_{\omega_L}$, and $\beta_q=-i{\bf g}(\omega_L)\sqrt{q}\cdot {\bf d}_{q\omega_L}$. Here, ${\bf d}_{\omega}=\int_0^T f(t)e^{i\omega t}\langle\hat{\bf d}_H(t)\rangle$, is a Fourier transform of the time averaged semiclassical dipole moment (Methods) weighted by the pulse envelope $f(t)$. Assuming that $N$ atoms contribute to HHG coherently in a phase matched way, we obtain that the final state of the fundamental and harmonic fields, after the pulse being coherent, is $|\alpha_L +\delta\alpha_L\rangle$ and $|\beta_q\rangle$ (Fig. 1) respectively, with  
	\begin{eqnarray}
		\delta\alpha_L &=&-iN{\bf g}(\omega_L)\cdot {\bf d}_{\omega_L}, \\
		\beta_q &=&-iN{\bf g}(\omega_L)\sqrt{q}\cdot {\bf d}_{q\omega_L}.
	\end{eqnarray}
The above, rigorous and analytic expressions, constitute one of the main results of this work as they provide a direct solution to the problem concerning the quantum nature of light in strong-field laser-atom interaction, and the key parameters that can be used to control the properties of the light states exiting the atomic medium.

\begin{figure*}[p]
\begin{center}
\includegraphics[width=15 cm]{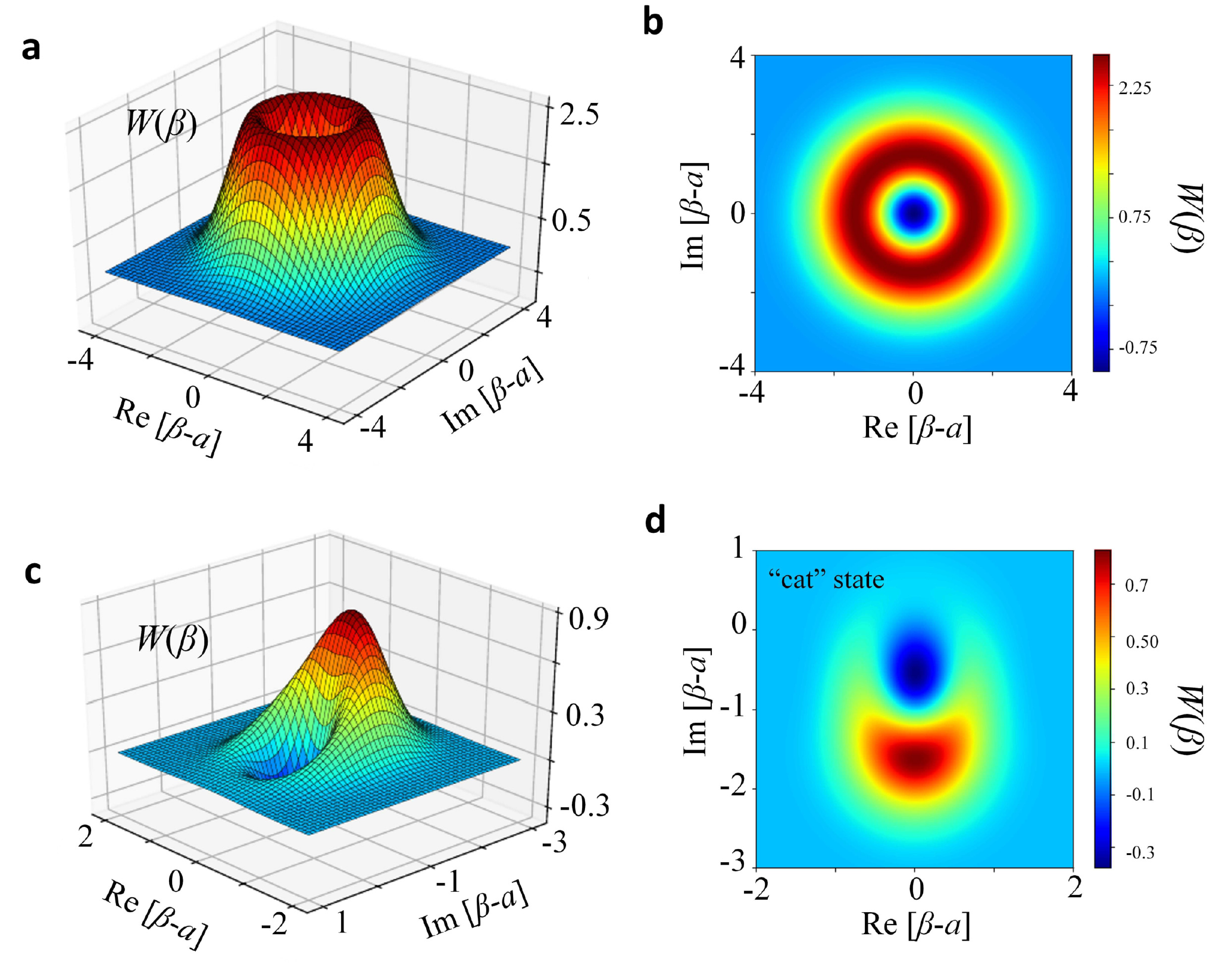}
\end{center}
\vspace{-8mm}
\caption{\textbf {Calculated Wigner functions of a Schr\"{o}dinger ``kitten'' and a ``cat'' state.} The Wigner functions $W(\beta)$ are plotted according to the terminology of ref.\cite{Schleich}. $\beta$ is a variable such that $Re[\beta-\alpha]\equiv x$ and $Im[\beta-\alpha]\equiv p$. $\it x$, $\it p$ are the values of the quadrature field operators $\hat{x}=(\hat{a}+\hat{a}^\dagger)/\sqrt{2}$ and $\hat{p}=(\hat{a}-\hat{a}^\dagger)/i\sqrt{2}$. $\delta\alpha$ represents the amplitude shift of the initial coherent state $|\alpha\rangle$ and $\zeta=\langle\alpha|\alpha +\delta\alpha\rangle$ reflects the coupling between the initial coherent state and the shifted one $|\alpha +\delta\alpha\rangle$. ({\bf a}) Wigner function of a Schr\"{o}dinger ``kitten'' state, corresponding to a coherently shifted first Fock state for small  $\delta \alpha $ where $|\zeta| \approx 1$. ({\bf b}) Projection of $W(\beta)$ shown in (a) on the $(Re[\beta-\alpha], Im[\beta-\alpha])$ plane. ({\bf c}) Wigner function of a genuine Schr\"{o}dinger ``cat'' state for $|\delta \alpha| = 1.5 $ (comparable with  $|\alpha| = 2$) where $|\zeta| \approx 0.32$. The Wigner depicts a ring structure with a maximum $W_{max}\approx 0.8$ at $(x, p) \approx (0, -1.7)$ and a negative minimum $W_{min}\approx -0.3$ at $(x, p) \approx (0, -0.7)$. The contrast $C=2(W_{max}-W_{min})/(W_{max}+W_{min})\approx 4.4$. It is noted that the exact shape of the Wigner function depends on the parameter $\zeta$. Here, we have used a $\zeta$ which provides a function reasonably close to the experimental results. ({\bf d}) Projection of $W(\beta)$ shown in (c) on the $(Re[\beta-\alpha], Im[\beta-\alpha])$ plane.}

\vspace{3mm}

\begin{center}
\includegraphics[width=15 cm]{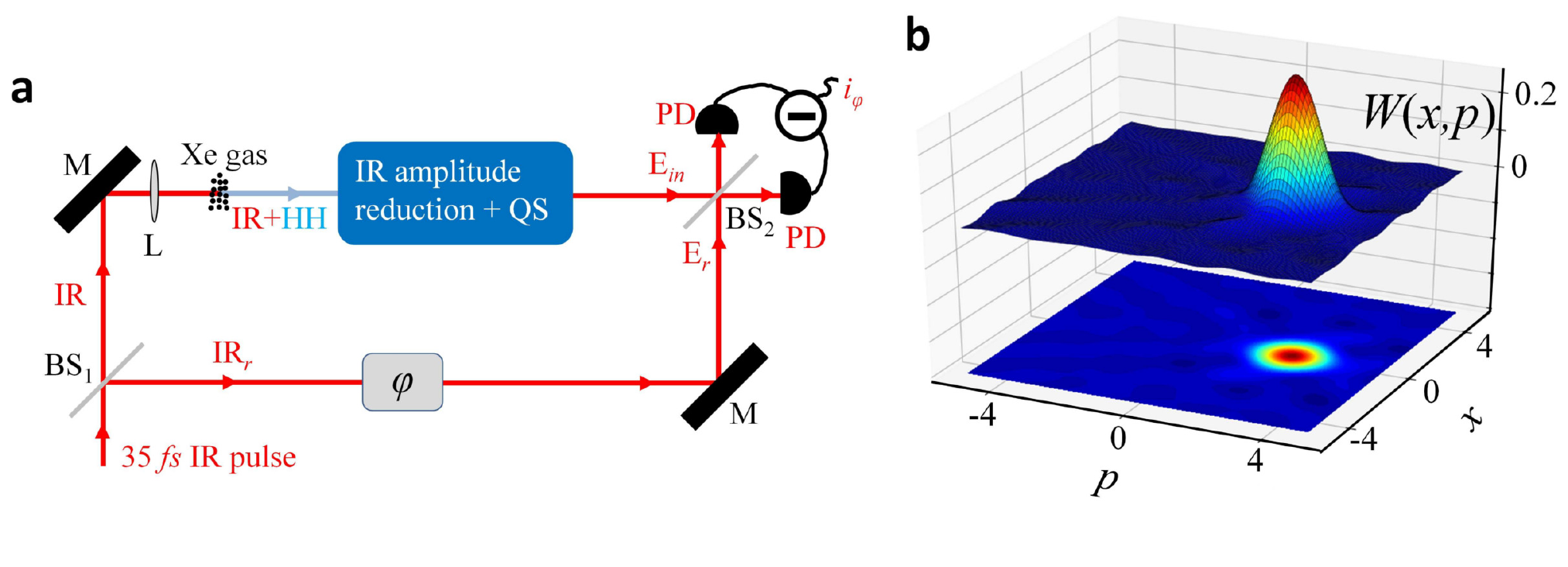}
\end{center}
\vspace{-8mm}
\caption{\textbf {Experimental approach and Wigner function of the coherent state of the laser.} ({\bf a}) Experimental set-up. BS$_{1}$ is an 80\%:20\% (transmission{:}reflection) IR beam separator. IR is the beam passing through BS$_{1}$ and focused by a lens (L) onto a Xenon gas jet where high harmonics (HH), with $q \leq 23$, were generated. The mean photon number of the IR field exiting the medium was significantly reduced to the level of few photons per pulse. The Quantum Spectrometer (QS), using the integrated photon number of the harmonics with $q \geq 11$ (i.e. harmonics lying on the plateau and cut-off region of the harmonic spectrum), selects for each laser shot only the IR photons related to the HHG process. IR$_{r}$ is the beam reflected by the BS$_{1}$. M are IR plane mirrors. BS$_{2}$ is an 50\%:50\% IR beam splitter. PD are identical IR photodiodes. E$_{r}$ is the local oscillator laser field of the reference beam and E$_{in}$ is the light field to be characterized. $\varphi$ is the phase shift introduced between the E$_{r}$ and E$_{in}$ fields. $i_{\phi}$ is the photocurrent difference used for the reconstruction of the Wigner function. ({\bf b}) Wigner function $W(x, p)$ of the coherent state (with $\langle{n}\rangle\approx5.27\pm0.04$) of the driving IR laser field measured when the HHG process was switched--off (i.e. Xenon gas jet and QS were switched--off). $\it x$, $\it p$ are the values of the quadrature field operators $\hat{x}=(\hat{a}+\hat{a}^\dagger)/\sqrt{2}$ and $\hat{p}=(\hat{a}-\hat{a}^\dagger)/i\sqrt{2}$. The error of the amplitude of $W(x, p)$ is $\pm0.001$.}
\end{figure*}

The consequences of this result towards the generation of non-classical light states is coming from the projection of the fundamental mode on its part corresponding to the HHG, which can be achieved experimentally by utilizing the QS approach\cite{Paris2,Paris3}. In order to show the result of this action, we use the conditions of the experiment presented below, where the QT of the fundamental mode, after its projection to HHG, is performed in two steps (Fig. 1). The first consists in reducing the amplitude of this mode by a factor $\cos(r)$. In this way, $\alpha_L \to  \cos(r)\alpha_L=\alpha$ and $\delta\alpha_L \to  \cos(r)\delta\alpha_L=\delta\alpha$. The second step corresponds to condition on HHG, i.e. post-selection of the part of the state that includes at least one harmonic photon. Mathematically it means that
	$|\alpha +\delta\alpha\rangle \to (1-|\alpha\rangle\langle \alpha|) |\alpha~+~\delta\alpha\rangle \nonumber =|\alpha +\delta\alpha\rangle -\langle\alpha|\alpha +\delta\alpha\rangle |\alpha\rangle$, that is, the HHG conditioning allow us to project the final coherent state onto the part of the fundamental field that has been affected by the HHG process, i.e., everything that was not in the initial state $|\alpha\rangle$. 
	This is the action that shows the coupling of the initial coherent state with the shifted one (via the matrix element $\zeta=\langle\alpha|\alpha +\delta\alpha\rangle$⟩) and creates the superposition of the shifted coherent state with the initial state, weighted by the coupling factor $\zeta$. It is elementary to see that the state is a superposition  of the two coherent states, so it is a Schr\"{o}dinger ``cat'' state. However, if the two coherent states are very close to each other, this is a ``kitten'' rather than a genuine ``cat''. In fact, we obtain two limiting cases:
\begin{enumerate}
\item[I] If $|\zeta|\approx 1$, then the post-selected state $|\phi_{\rm post}\rangle$, conditioned on HHG, is nevertheless quite non-classical corresponding to a shifted Fock state,
	$|\phi_{\rm post}\rangle = (\hat a^\dagger-\alpha^*)|\alpha\rangle$,
	having a Wigner function $W(\beta) = (4|\beta-\alpha|^2 -1) e^{-|\beta-\alpha|^2/2}$ (Fig. 2 a, b). 
\item[II] If $0<|\zeta|< 1$  the state $|\phi_{\rm post}\rangle$ corresponds indeed to a ``cat'' state, depicting a Wigner function $W(\beta)= \allowbreak{} e^{-2|\beta - \alpha - \delta \alpha|^2} \allowbreak{} + \allowbreak{}  e^{-|\delta \alpha|^2} \cdot e^{-2 |\beta - \alpha|^2}- \allowbreak{} \big(e^{2(\beta - \alpha)\delta \alpha^*}+ e^{2(\beta - \alpha)^*\delta \alpha}\big) \cdot e^{-|\delta \alpha|^2} \cdot e^{-2 |\beta - \alpha|^2}$ (Fig.~2 c, d) with negative regions \cite{Schleich}.
\end{enumerate}

\begin{figure*}[t]
\begin{center}
\includegraphics[width=15 cm]{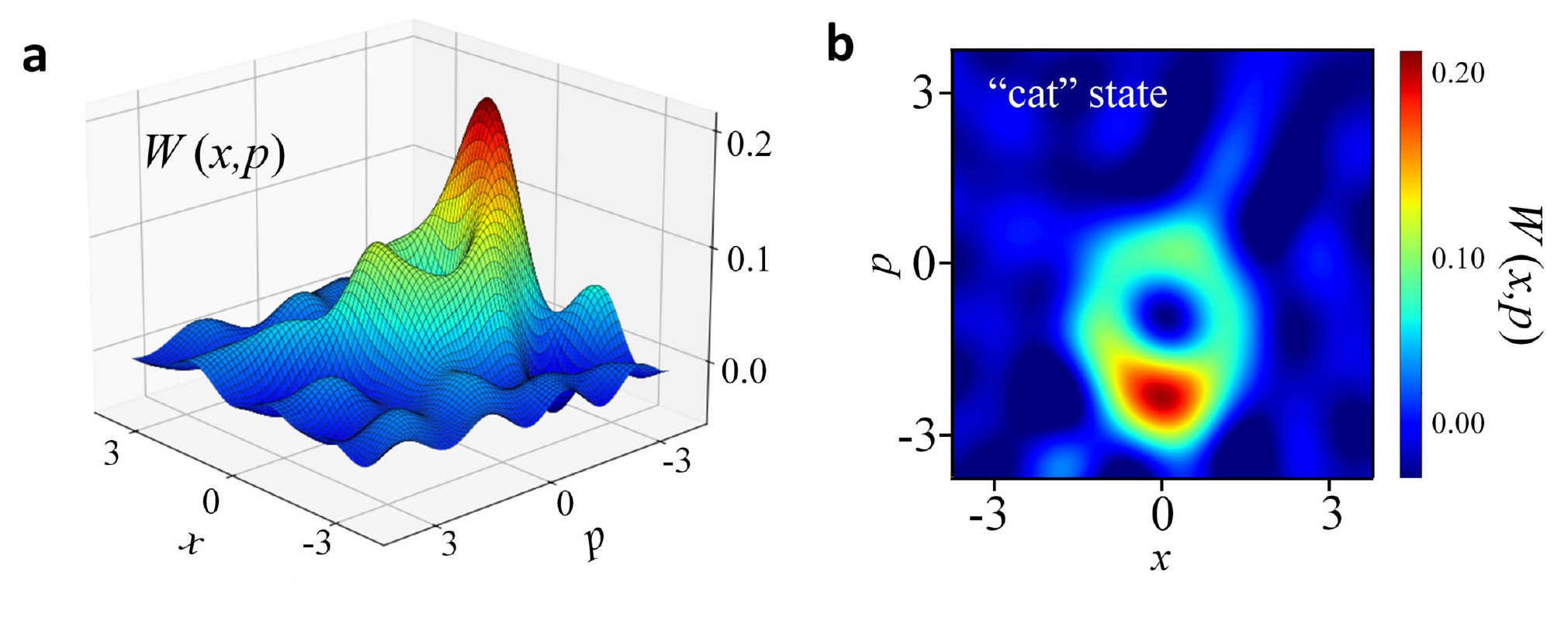}
\end{center}
\caption{\textbf {Measurement of the Wigner function of the genuine Schr\"{o}dinger ``cat'' state.}  ({\bf a}) Measured Wigner function $W(x, p)$ of the IR field (with $\langle{n}\rangle\approx1.98\pm0.04$) when the HHG process and the conditioning to the HHG, were switched--on (i.e. Xenon gas jet and QS switched--on). $\it x$, $\it p$ are the values of the quadrature field operators $\hat{x}=(\hat{a}+\hat{a}^\dagger)/\sqrt{2}$ and $\hat{p}=(\hat{a}-\hat{a}^\dagger)/i\sqrt{2}$. The error of the amplitude of $W(x, p)$ is $\pm0.002$. In agreement with the theoretical predictions, the function clearly shows a Schr\"{o}dinger ``cat'' state. It depicts a ring structure with a maximum $W_{max}\approx 0.21$ at $(x, p) \approx (0, -1.9)$ and a negative minimum $W_{min}\approx -0.004$ at $(x, p) \approx (0, -0.8)$. ({\bf b}) Contour color plot of the projected Wigner function on the $(x, p)$ plane.}
\end{figure*}

The experimental results presented here suggest the generation of a ``cat'' state (i.e case II). Since $\alpha_L\simeq 10^6$, that would require that $N{\bf g}(\omega_L)\cdot {\bf d}_{\omega_L}\simeq 10^6$. We expect the atomic dipole moment to be of the order of $e$ times the electron excursion amplitude, that is 10-100 a.u.. This would imply that the number of atoms participating in the coherent, phase matched HHG is $N\simeq 10^{12}-10^{13}$, which is consistent, within the large uncertainty, with the experimental estimations (Supplementary Sections 1, 2). The experiment has been performed using the set-up shown in Fig. 3a (for more details see Supplementary Section 1). The arrangement consists of an interaction area where the harmonics are generated, and the QT, QS approaches. The QS was used to project the fundamental mode on its part corresponding to the harmonic generation (Supplementary Section 3), and the QT to characterize the quantum state of the light field exiting the atomic medium (Supplementary Section 4). In the 1st branch of the interferometer, the linearly polarized $\approx$ 35 fs infrared (IR) laser pulse, after passing through a beam separator (BS$_1$), was focused with an intensity $\approx 8\times 10^{13}$ W/cm$^2$ into Xenon atoms. It is here where the strong-field laser-atom interaction and HHG process takes place and high--harmonics with $q \leq 23$ were generated (Supplementary Section 1). The mean photon number of the IR field exiting the medium was reduced to the level of few photons per pulse, with the QS to select, for each laser shot, only the IR photons related to the HHG. The light field (E$_{in}$) after QS was spatiotemporally overlapped in a beam splitter (BS$_2$) with the local oscillator laser field (E$_{r}$) coming from the 2nd branch of the interferometer, which contains a translation stage that introduces a phase shift $\varphi$ between the E$_{r}$ and E$_{in}$ fields. The power of the E$_{r}$ was about $\sim 10^7$ higher than the E$_{in}$. The interfering fields after BS$_2$ were recorded by a balanced detector which provides at each value of $\varphi$ the photocurrent difference $i_{\varphi}$. These values correspond to the measurement of the electric field quadrature operator, and have been used for the reconstruction of the Wigner function according to the well known methodology described in refs.\cite{Knight,Lvovsky,Breitenbach} (for details see Supplementary Section 4). As expected, when the HHG process was switched--off, the state of the driving IR laser field is coherent depicting a Wigner function with Gaussian distribution (Fig. 3b). When the HHG process was switched--on, the state of the IR light field (conditioned to the HHG), in agreement with the theoretical predictions, shows a ``cat'' state (Fig. 4a, b) (for details see Supplementary Sections~4,~5).

Finally, we note that the harmonic emission and the shift of the coherent state of the fundamental mode are correlated. Indeed, the whole process is governed by a ``wave packet'' mode described by a creation operator ${\hat B}^\dagger(t)\propto {\hat a}^\dagger e^{i\omega_L t} + \sum_{q=3}^\mathrm{cutoff} \sqrt{q}\hat b_q^\dagger e^{iq\omega_L t}$ (Methods). The global coherent state is a product state, but only the effective mode described by this equation is excited/shifted; all other orthogonal modes remain unaffected. The parts corresponding to the fundamental and the harmonics oscillate at very different frequencies, and thus the observation of such correlations would require the incorporation of approaches that provide attosecond time resolution\cite{Krausz2009}, into the QT scheme.

In conclusion, we demonstrate the quantum nature of the electromagnetic field in intense laser-atom interactions, and a method for producing optical Schr\"{o}dinger ``cat'' states. Using the HHG process, we show that the quantum states of the fundamental and the harmonic modes after the interaction are coherent, with the fundamental to be shifted due to quantum effects. When the latter is conditioned onto HHG, it results in a superposition of two coherent states which interpolates between a ``kitten'' and a ``cat'' state. This was experimentally confirmed by measuring the Wigner function of the ``cat'' state. The method is applicable for the production of large optical ``cat'' states and in a large variety of intense laser-matter interactions\cite{Nayak}.






\section{Methods}
\noindent\textbf{Theoretical approach.}
Of course, the theoretical approach relies on several simplifying assumptions, but all of them can be avoided using less restrictive approximations. For instance, the reason why we can achieve these results is because we condition our analysis on the electrons being in or back in the ground state, ``ignoring'' in a sense the electrons being ionized. Also, in order to describe a laser pulse with a given envelope, one needs a continuum of photon modes contributing to this wave packet. We avoid this complexity by multiplying the atom-field interaction Hamiltonian by an envelope  function  $f(t)$, in some analogy to scattering theory in many body quantum field theories. Our starting point is the time-dependent Schr\"odinger equation,
\begin{equation}
i\hbar\frac{ \partial}{ \partial t} | \tilde{\Psi}(t) \rangle=\hat{H} | \tilde{\Psi}(t) \rangle, 
\label{Eq:SE}
\end{equation}
where the Hamiltonian, $\hat{H}$, describes the laser-target system in single active electron approximation (SAE), and is the sum of three terms, i.e. $\hat{H} = \hat{H_0}+\hat{U}  + \hat{H_f} \label{Eq:H}$, where $\hat{H}_0 = \frac{\hat{\textbf{p}}^2}{2m} + V(\hat{\textbf{r}})$ is the laser-free Hamiltonian of the atomic or molecular system
with $V(\hat{\textbf{r}})$ being the effective SAE atomic or molecular potential, $m$ the electron mass; $\hat{U}=-e \hat{\bf E}\cdot \hat{\bf r}$ the  dipole coupling, which describes the interaction of the atomic or molecular system with the laser radiation, written in the length gauge and under the dipole approximation; finally, $H_f = \int d \omega \ \hbar \omega \hat{a}^\dagger_\omega \hat{a}_\omega$ is the electromagnetic (EM) field Hamiltonian containing all the frequency modes.

In principle, to describe laser/harmonic  pulses of finite duration and spatial extension, the full continuum spectrum of the EM field must be  considered, as stated in our original definition of $H_f$. Here, we simplify the Hamiltionian to a sum of effective discrete modes. The free EM field Hamiltonian reduces in our case to
\begin{eqnarray}
\hat{H}_f&=&\hbar \omega \hat a^\dagger \hat a + \sum_q^\mathrm{cutoff} \hbar \omega q\hat b_q^\dagger \hat b_q,
\end{eqnarray}
where $\hat a^\dagger$,  $\hat a$, $\hat b_q^\dagger$, $\hat b_q$ are creation (annihilation) operators of the laser and harmonic modes, respectively.  To account for finite pulse duration we model the electric field operator as
\begin{equation}
{\hat{\bf E}}= -i\hbar {\bf g}(\omega_L)f(t)\left [ (\hat{a}^\dag -\hat{a}) + \sum_3^\mathrm{cutoff}\sqrt{q}(\hat{b}_q^\dag - \hat{b}_q)\right].
\label{efield}
\end{equation}
In  Eq. (\ref{efield}) the dimensionless function $0\le f(t)\le 1$ describes the envelope of the laser pulse normalized to one at maximum. We denote the effective coefficient entering into the expansion of the electric field into the modes by ${\bf g}(\omega)\propto \sqrt{\omega/V_{eff}}$, where $V_{eff}$ is the effective quantization volume$^{33, 34}$. $e{\bf g}(\omega)$ encodes information about the polarization of the modes, has dimension [1/m/sec], and for the typical frequencies used in strong field physics is very small, on the order of $10^{-8}$ in a. u..

Equation (\ref{Eq:SE}) needs to be solved starting from the initial condition $| \tilde{\Psi}(0) \rangle = |g, \alpha_L, \Omega_H\rangle$, i.e. for the electron initially being in its ground state $|g\rangle$, the laser in a coherent state $|\alpha_L\rangle$, and the harmonics in the vacuum state $|\Omega_H\rangle$. To this aim we write $|\tilde{\Psi}(t)\rangle = \exp[-i\hat{H}_f t/\hbar]D(\alpha_L) |\Psi(t)\rangle$, where $D(\alpha_L)$ is the Glauber's shift operator creating a coherent state $|\alpha_L\rangle$ from the vacuum of the laser mode, $|\Omega_L\rangle$. The second unitary operator transforms to the interaction  picture with respect to the EM field. With these transformations, the initial state of the system is now described by $|\Psi(0)\rangle = |g, \Omega_L, \Omega_H\rangle$ and the electric field part of the Hamiltonian shown in Eq.~(\ref{Eq:SE}) becomes time-dependent and gains an extra factor describing the behaviour of the ``classical'' field, i.e., the mean value $\langle \alpha_L |{\bf \hat{E}}(t) |\alpha_L \rangle$. More explicitly, our Schr\"{o}dinger equation now reads
\begin{equation}
i\hbar\frac{\partial}{ \partial t} |\Psi(t) \rangle=
\left[\hat{H}_{sc} -e\hat{\bf E}_Q(t)\cdot \hat{\bf r}\right]| \Psi(t) \rangle,
\end{equation}
where $\hat{H}_{sc}(t)=\hat{H}_0-e {\bf E}_{L}(t)\cdot \hat{\bf r}$, and ${\bf E}_{L}(t)=-i\hbar {|\bf g}(\omega_L)|f(t)[\alpha^*_Le^{i\omega_l t} -\alpha_L e^{-i\omega_l t} ]$ the ``classical'' electric field of the laser pulse. The quantum correction is
\begin{align}\label{Qu:fluc}
\hat{\bf E}_Q(t)
&=-i\hbar {\bf g}(\omega_L)f(t)\bigg[\hat a^\dagger e^{i\omega_L t} -\hat a e^{-i\omega_L t} 
\nonumber\\
&\qquad \quad +
\sum_{q=3}^\mathrm{cutoff} \sqrt{q}[\hat b_q^\dagger e^{iq\omega_L t} -\hat b_q e^{-iq\omega_L t} ]\bigg].
\end{align}
		
The next step is to go to the interaction picture with respect to
$\hat{H}_{sc}(t)$, something that we achieve with the following transformation
\begin{equation}
\label{Semic:transf}
|\Psi(t) \rangle ={\cal T}\exp\Big[-i\int_0^t \hat{H}_{sc}(t')dt'/\hbar\Big]|\psi(t) \rangle,    
\end{equation}
where $\cal T$ denotes the time ordered product.  Then we obtain:
\begin{equation}
i\hbar\frac{\partial}{ \partial t} |\psi(t) \rangle=
-e\hat{\bf E}_Q(t)\cdot \hat{\bf r}_H(t)|\psi(t) \rangle.
\label{eqpsi}
\end{equation}
		
The transformation depicted in Eq.~(\ref{Semic:transf}) does not alter the initial condition for the system, but introduces in our Schr\"{o}dinger equation the dynamics driven by the semiclassical term $H_{sc}(t)$ through the time-dependent dipole operator $e\hat{\bf r}_H(t)$, written now in the Heisenberg picture with respect to $H_{sc}(t)$. Due to this evolution, the electron might be ionized in the continuum (above threshold ionization process, ATI), or (hardly) remains in some bound excited state. Most of the physics relevant for HHG happens in the ground state: whatever remains there, remains there; whatever recombines to   it, is related to a harmonic emission. Therefore, it makes sense to condition Eq. (\ref{eqpsi})
on the ground state, i.e. to consider $|\phi(t)\rangle=\langle g|\psi(t)\rangle$, which fulfills Eq.~(\ref{eqphi}) of the main text, i.e. 
\begin{equation}
i\hbar\frac{\partial}{ \partial t} |\phi(t) \rangle=
-\hat{\bf E}_Q(t)\cdot \langle\hat{\bf d}_H(t)\rangle|\phi(t) \rangle,
\label{eqphi:SM}
\end{equation}
where $\hat{\bf E}_Q(t)$ is the quantum fluctuating part of the laser fields described in Eq.~(\ref{Qu:fluc}), and $\langle\hat{\bf d}_H(t)\rangle$ is the quantum averaged time-dependent dipole moment over the ground state, i.e., $\langle g| e \hat{\bf r}_H(t)|g \rangle$, that can be efficiently calculated solving the TDSE, or even easier using the strong field approximation (SFA) \cite{Lewenstein1994, Symphony}. Also, the effective Hamiltonian $\hat{H}_Q(t)= -\hat{\bf E}_Q(t)\cdot \langle\hat{\bf d}_H(t)\rangle$ is a linear form of photon creation and annihilation operators. Thus, the unitary evolution operator is an exponent of a linear form of creation and annihilation operators, and thus when acting on coherent states, it will shift them. One can obviously calculate exactly the evolution operator corresponding to $\hat{H}_Q(t)$. Since $[\hat{H}_Q(t), \hat{H}_Q(t')]={\rm c(t-t')}$ is simply a complex number, this is trivially achieved applying multiple times the Baker-Campbell-Hausdorff formula.The final result (up to an uninteresting prefactor) at time $T$ after termination of the laser pulse, and after returning to the ``laboratory'' frame, is shown in Eq.~(2) of the main text of the manuscript. We note, that the present theoretical approach is consistent with results provided by the semiclassical analysis. This can be seen in Eqs. (3) and (4), where the amplitude shift of the fundamental and the amplitude of the harmonics scale directly with the spectral amplitude.

We also note that the photon number distributions of the shifted Fock state (case I where $|\zeta|\approx 1$) and genuine Schr\"{o}dinger ``cat'' state (case II where $0<|\zeta|< 1$) are $p_n=|n/\alpha-\alpha^*|^2|\alpha|^{2n}\exp(-|\alpha|^2)/n!$ and 
$p_n=|(\alpha+\delta\alpha)^{n}\exp(-|\alpha+\delta\alpha|^2/2)-
\langle\alpha|\alpha +\delta\alpha\rangle\alpha^{n}\exp(-|\alpha|^2/2)|^2/n!$, respectively.

Regarding the connection of the projector $1 - |\alpha\rangle \langle\alpha|$ with the QS approach: As we have mentioned in the main text of the manuscript, the harmonic emission and the shift of the fundamental are correlated. The whole process is governed by a "wavepacket" mode described by the creation operator ${\hat B}^\dagger(t)\propto {\hat a}^\dagger e^{i\omega_L t} + \sum_{q=3}^\mathrm{cutoff} \sqrt{q}\hat b_q^\dagger e^{iq\omega_L t}$. Thus, when we experimentally introduce the HHG conditioning via the QS/QT approach, from the theoretical point of view we are looking at excitations generated by the above creation operator. Hence, if there is a non--vanishing HHG signal we know that we have obtained a coherent shift $\delta \alpha \neq 0$, and on the other hand if the HHG signal vanishes then $\delta \alpha = 0$. In the presented work we are interested in the first case as we do measure a HHG signal, and this is effectively described by the projector $1 - |\alpha\rangle \langle\alpha|$.

\section{Data availability}
All other data that support the plots within this paper are available from the corresponding authors on reasonable request. Source data are provided with this paper.

\section{Code availability}
The codes used in this study are available from the corresponding authors upon request.

\balance
	
\section{Acknowledgements}
We dedicate this work to the memory of Roy J. Glauber, the inventor of coherent states. We thank Jens Biegert, Ido Kaminer, and Pascal Sali\`eres for enlightening discussions. We also thank I. Liontos, E. Skantzakis, Prof. S. Karsch from Max Plank Institute for Quantum Optics for his assistance on maintaining the performance of the Ti:Sa laser system and N. Pappadakis for his contribution on the development of the data acquisition and data analysis system. M.L. group acknowledges the European Research Council (ERC AdG) NOQIA, Spanish Ministry MINECO and State Research Agency AEI (FIDEUA PID2019-106901GB-I00/10.13039 / 501100011033, SEVERO OCHOA No. SEV-2015-0522 and CEX2019-000910-S, FPI), European Social Fund, Fundació Cellex, Fundació Mir-Puig, Generalitat de Catalunya (AGAUR Grant No. 2017 SGR 1341, CERCA program, QuantumCAT$\_$U16-011424 , co-funded by ERDF Operational Program of Catalonia 2014-2020), MINECO-EU QUANTERA MAQS (funded by State Research Agency (AEI) PCI2019-111828-2 / 10.13039/501100011033), EU Horizon 2020 FET-OPEN OPTOLogic (Grant No 899794), and the National Science Centre, Poland-Symfonia Grant No. 2016/20/W/ST4/00314. M. F. C. acknowledges the Grantová agentura Ceské Republiky (GACR Grant 20-24805J). J.R-D has received funding from the Secretaria d'Universitats i Recerca del Departament d'Empresa i Coneixement de la Generalitat de Catalunya, as well as the European Social Fund (L'FSE inverteix en el teu futur)--FEDER. P.T. group acknowledges LASERLABEUROPE (H2020-EU.1.4.1.2 Grant ID 654148), FORTH Synergy Grant AgiIDA (Grand No. 00133), the EU's H2020 framework programme for research and innovation under the NFFA-Europe-Pilot project (Grant No. 101007417), the “HELLAS-CH” (MIS Grant No. 5002735) [which is implemented under the “Action for Strengthening Research and Innovation Infrastructures,” funded by the Operational Program “Competitiveness, Entrepreneurship and Innovation” (NSRF 2014–2020) and co-financed by Greece and the European Union (European Regional Development Fund)], and the European Union’s Horizon 2020 research. ELI-ALPS is supported by the European Union and co-financed by the European Regional Development Fund (GINOP Grant No. 2.3.6-15-2015-00001).

\section{Author contributions}
M.L.: Supervised the theoretical part of the work; M.F.C., J.R.-D., E.P.: equally contributed to the manuscript preparation and the development of the theoretical approach; P.S.: Contributed to the theoretical calculations; Th.L.: contributed in the experimental runs and data analysis; P.T.: Supervised the experimental part of the work.

\hfill

\newpage

\renewcommand\thesection{S\arabic{section}}
\setcounter{section}{0}
\titleformat{\section}[block]{\bfseries\upshape\sffamily\boldmath}{\thesection}{0.5em}{}

\renewcommand\thefigure{S\arabic{figure}}
\setcounter{figure}{0}

\renewcommand\theequation{S\arabic{equation}}
\setcounter{equation}{0}

\twocolumn[
\begin{@twocolumnfalse}
{
\raggedright\huge\bfseries\upshape\sffamily\boldmath
\color[rgb]{0,0,0.8}
Supplementary Information
}
\vspace{5mm}
\end{@twocolumnfalse}
]

\section{Experimental set-up}

The optical layout of the set-up is shown in the Fig. S1a. The measurement was performed using a linearly polarized $\approx$ 35 fs Ti:Sapphire laser pulse of $\lambda\approx$ 800 nm carrier wavelength and an interferometer. The whole system was operating at 0.5 kHz repetition rate recording data for each laser shot. The IR laser beam was separated into the branches of the interferometer by a beam separator BS$_1$, having 80\% transmission and 20\% reflection. The reflected by the BS$_1$ IR beam (in the $2^{nd}$ branch of the interferometer) serves as a reference beam of the Quantum Tomography method and for measuring (by means of PD$_{0}$) the shot-to-shot energy fluctuations of the driving field. In the 1$^{st}$ branch of the interferometer, the IR beam (IR$_t$) was focused by means of a 15 cm focal length lens (L1) into a Xenon pulsed gas jet, where the strong-field laser-atom interaction and HHG process takes place. The generated harmonics, after a reflection by a multilayer infrared-antireflection coating plane mirror (HS) placed at grazing incidence angle, was passing through a 150 nm thick Aluminum filter, which selects all the harmonics with $q \geq 11$. The harmonic spectrum was measured by means of a conventional XUV--spectrometer (not shown in Fig. S1a) and the photon number of the XUV radiation was measured by means of a calibrated XUV detector PD$_{HH}$ and used by the QS. The harmonic signal, and thus the phase matching conditions, was optimized by varying the gas density $\varrho_{Xe}$ and the laser intensity $I_{t}$ in the interaction region. At the optimum conditions the presence of strong plasma effects, are minimized and considered negligible. Additionally, we emphasize that the QS approach allows the characterization of the quantum state of the IR field which is relevant only with the HHG i.e. potential effects irrelevant to HHG, caused by the propagation in the medium, do not contribute. The optimum harmonic generation conditions were achieved for $\varrho_{Xe}\sim 3\times 10^{18}$ atoms per cm$^3$ and $I_{t} \approx 8 \times 10^{13}$ Watt per cm$^2$. The value of $\varrho_{Xe}$ was obtained taking into account the characteristics of the piezo pulsed gas nozzle\cite {Altucci} and the experimental conditions that have been used in the present study. The Xe backing pressure of the pulsed nozzle was $\approx 3$ bar and the IR beam was focused $\approx$ 100 $\mu$m to $\approx$ 300 $\mu$m below the nozzle orifice (of $\approx 1$ mm diameter), which results to a medium length $L_{med}$ in the range of 1.2 mm to 1.4 mm. The $I_{t}$ was obtained taking into account the measured focal spot diameter ($\approx 45$ $\mu$m), pulse duration ($\approx 35$ fs) and the mean photon number ($\approx 2\times10^{14}$ photons per pulse) of the driving IR field before the interaction. This intensity value was confirmed by measuring the energy of the cut-off harmonic in the spectrum (shown in Fig. S1b), which was found to be at $\approx 27$ eV. This is in agreement with the value given by the cut-off law $E_\mathrm{cutoff} \approx IP_{Xe} + 3.17 U_{p} \approx 28$ eV \cite {Lewenstein1994} (where $IP_{Xe}=12.13$ eV is the ionization potential of Xe and $U_{p}$ is the electron ponderomotive energy). Also, in this configuration the confocal parameter of the fundamental beam is longer than the medium length and thus the intensity of the laser beam along the propagation in Xenon gas was considered constant. A portion ($\approx 10$\%) of the IR field exiting the Xenon gas was reflected by the IR beam separator BS$_3$ towards IR photodiode PD$_{out}$ which, in order to avoid saturation, was placed after a neutral density filter (F). The photocurrent signals $i_{HH}$, $i_{0}$, $i_{out}$  of PD$_{HH}$, PD$_{0}$ and PD$_{out}$ were used by the QS to disentangle the high-harmonic generation process from all other processes induced by the interaction (see the Quantum Spectrometer Section 3). The IR field after BS$_3$ was collimated by a concave lens (L$_{2}$) while the mean photon number of the IR field (E$_{in}$) before reaching the balanced detector of the QT (see the Quantum Tomography Section 4), was significantly reduced by means of neutral density filters (F$_{in}$). For the measurements shown in Fig. 4 of the main text of the manuscript the mean photon number was reduced to the level of few photons per pulse, with the QS to select, for each laser shot, only the IR photons related to the HHG.
	
The E$_{in}$ field was spatiotemporally overlapped on a 50:50 beam splitter (BS$_{2}$) with an unaffected by the interaction local oscillator laser field (E$_{r}$) coming from the 2$^{nd}$ branch of the interferometer which consists of a delay stage that introduces a controllable delay $\Delta\tau$ (phase shift $\varphi$) between the E$_{r}$ and E$_{in}$ fields. The delay stage was equipped with a rough precision steeper motor (supporting a displacement down to $\approx$ 20 nm per step) that was used to temporally overlap and find the zero delay ($\Delta\tau\approx 0$) of the E$_{r}$ and E$_{in}$ fields, and a high precision piezo based stage (supporting a displacement down to $\approx$ 2 nm per step) that has been used to record the field quadrature around $\Delta\tau\approx 0$.
	
\begin{figure*}[h!]
\includegraphics[width=17.5 cm]{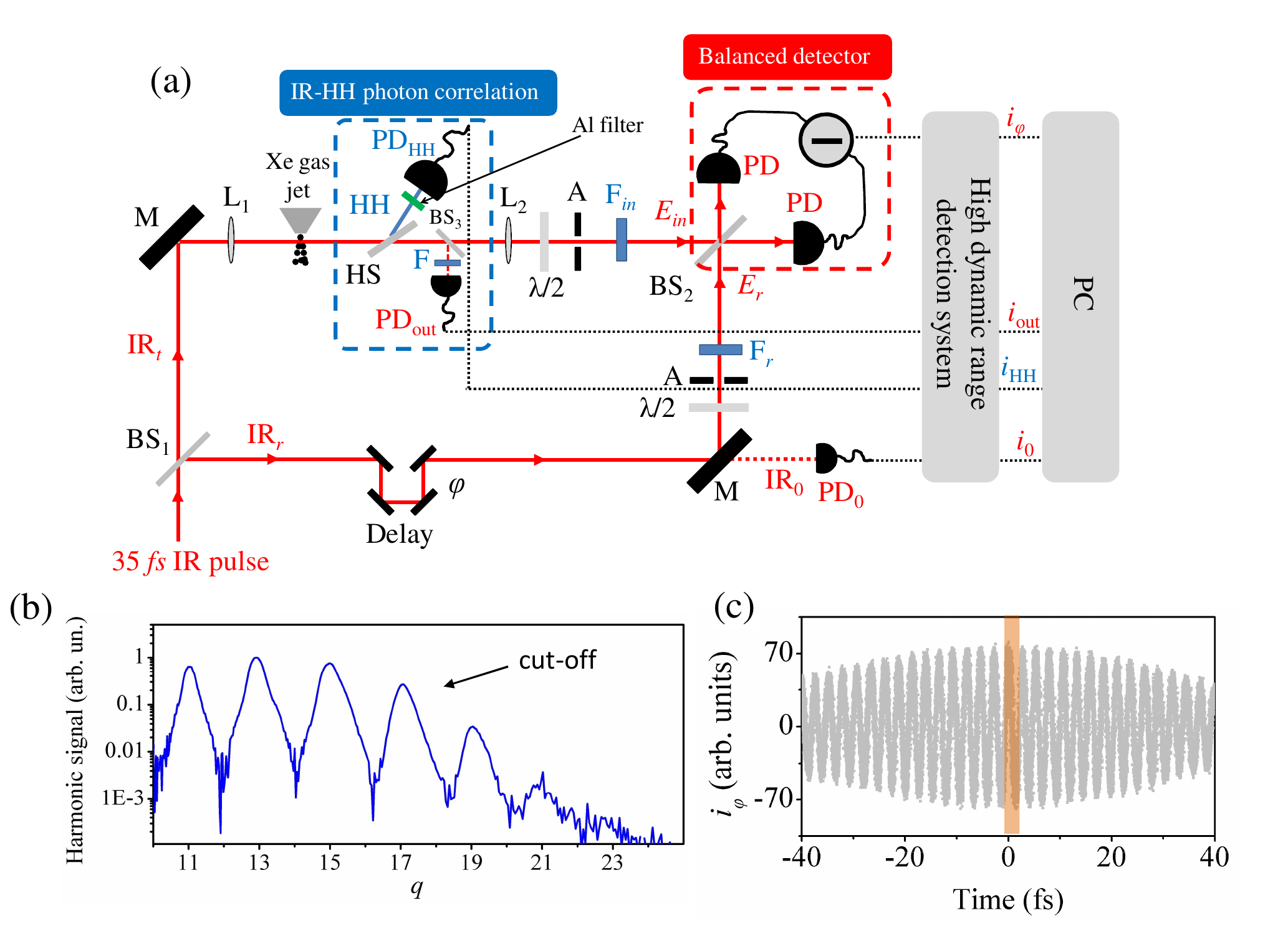}
\caption{\textbf {Experimental procedure.} (a) Optical layout of the experimental set-up. BS$_{1}$: 80\%:20\% (Transmission:Reflection) IR beam separator. IR$_{t}$: IR beam passing through BS$_{1}$. IR$_{r}$: IR beam reflected by BS$_{1}$. M: IR plane mirrors. L$_{1,2}$: Lens. HS: harmonic separator which reflects the high-harmonics and leaves the IR beam to pass through. HH: High-harmonics. BS$_{2,3}$: 50\%:50\% IR beam splitters. PD, PD$_{0}$, PD$_{out}$, PD$_{HH}$: IR and HH photodetectors. PD$_{0}$ is part of the quantum spectrometer approach. IR$_{0}$ is a small portion of the IR$_{r}$ beam which is used for measuring (by means of PD$_{0}$) the shot--to--shot energy fluctuations of the driving field.Just before PD$_{HH}$ a 150 nm thick Aluminum filter was placed in order to select the high harmonics lying in the plateau/cut-off spectral region of the generated harmonic spectrum. $\lambda/2$: Half-IR-wave plate. A: Aperture. F, F$_{in,r}$: Neutral density filters. E$_{r}$ is the local oscillator laser field of the reference beam and E$_{in}$ is the light field to be characterized. $\varphi$ is the phase shift introduced by the delay stage between the E$_{r}$ and E$_{in}$ fields. $i_{\phi}$, $i_{out}$, $i_{0}$, $i_{HH}$, are the photocurrent values recorded for each laser shot by a multichannel 16 bit high dynamic range boxcar integrator. For each shot the background electronic noise was recorded and subtracted by the corresponding photocurrent signals. The values were saved and analyzed by computer (PC) software. The values of $i_{out}$, $i_{0}$, $i_{HH}$ were used by the QS. (b) Spectrum of high harmonics recorded by PD$_{HH}$. $q$ is the harmonic order. The arrow depicts the position of the cut-off harmonic. (c) 1st--order autocorrelation trace recorded for $R\sim 10^{-2}$. The trace was used to temporally overlap and find the zero delay ($\Delta\tau\approx 0$) of the E$_{r}$ and E$_{in}$ fields (shaded area).}
\end{figure*}
	
The power of the E$_{r}$ was controlled by means of neutral density filters (F$_{r}$). The F$_{in}$ and F$_{r}$ were mounted on a motorized translation stages. As the BS$_{2}$ is not perfectly insensitive to the polarization, the energy balance between the beams at the output of the BS$_{2}$ was adjusted by fine tuning (by the same angle) the $\lambda/2$ plates placed in each branch of the interferometer. In order to eliminate any wavefront imperfections that may appear along the beam profile of the interference fields, a pair of $\approx$ 2 mm diameter apertures (A) were placed in the beam paths selecting only the central part of the beam profiles. The outgoing from the BS$_{2}$ interfering fields were detected by the diodes (PD) of a high bandwidth (from DC to 350 MHz), high subtraction efficiency and high quantum efficiency, balanced amplified differential photodetector, which provides at each value of $\varphi$ the signal difference. The photocurrent difference $i_{\varphi}$, as well as the photocurrent values of the IR and HH detectors ($i_{out}$, $i_{0}$, $i_{HH}$) in the QS, were simultaneously recorded for each laser shot by a multichannel 16 bit high dynamic range boxcar integrator. The photodiodes of the balanced detector were operating below their saturation threshold and its shot-noise power dependence in the linear regime. For each shot the background electronic noise was recorded and subtracted by the corresponding photocurrent signal by placing a second time-gate in the boxcar integrator in times significantly delayed compared to the arrival times of the photon signals. Care was taken to isolate/shield the interferometer, the electronics and the detectors from any environmental noises (mechanical/sound, electrical currents and photons). The experiment was remotely controlled in order to avoid introducing environmental noises during the measurements. A band block frequency filter operating in the range from $5 \times 10^{-3}$ Hz to 150 Hz eliminates the noise instabilities, introduced in the interferometer by the vacuum pumps and environmental mechanical noises, leaving practically unaffected the quantum noise fluctuations of the light state and the mean oscillation frequency of the photocurrent signal $i_{\varphi}$. To temporally overlap and find the zero delay ($\Delta\tau\approx 0$) of the E$_{r}$ and E$_{in}$ fields, their energy $R$ ratio was set to be in the range of $\sim 10^{-2}$ (Fig. S1c). Setting the stepper motor stage around $\Delta\tau\approx 0$ (shaded area in Fig. S1c), the characterization of the quantum state (and the measurements shown in the main text of the manuscript) of light, was achieved by setting the $R$ in the range of $\sim 10^{-7}$ and accumulating $\approx 10^5$ shots. The measurement was performed by moving the piezo from $\varphi = 0 - \varepsilon$ to  $\varphi =\pi + \varepsilon$ (where $\varepsilon \approx \pi / 8$). As no feedback control loop (that provides the absolute piezo position i.e. $\varepsilon$) was used, the calibration of the x-axis of the homodyne traces was achieved by corresponding to $\varphi = 0$ and $\varphi = \pi$ the maximum and minimum mean values of $i_{\varphi}$, respectively. The homodyne data were scaled according to the measured vacuum state quadrature noise. The Wigner functions have been reconstructed from the experimental data following the well known methodology of refs.~ \cite{Knight, Lvovsky, Breitenbach, Leonhardt} (see the Quantum Tomography Section 4).

\section{Estimation of the number of atoms and IR photons participating to the HHG process}

A rough estimation of the number of atoms participating in the HHG process can be obtained by considering that in realistic experimental conditions of an interaction volume in the range of $10^{-5}$ - $10^{-6}$ cm$^3$, and an atomic density $\sim 3\times10^{18}$ atoms per cm$^3$. In this case the number of atoms participating in the HHG process is in the order of $\sim 10^{13}$ atoms. A value of the IR photon number absorbed towards the harmonic emission can be roughly obtained by using the $\varrho_{Xe}$, the $L_{med}$ and the measured harmonic photon number $N_{HH}^{(meas.)}$. Taking into account the harmonic signal (integral over the plateau/cut-off harmonic spectrum) measured by the calibrated XUV detector, the $\approx 60$ \% transmission of the Aluminium filter placed before the XUV detector and the $\approx 50$\% reflectivity of the plane mirror we obtain a value of $N_{HH}^{(meas.)}\sim 7\times10^{7}$ photons per pulse.  The $N_{HH}^{(meas.)}$ results after the propagation of the harmonics in the medium, where the absorption effects significantly reduce (by a factor of $A$) the harmonic photon number exiting the medium. In order to obtain the total harmonic photon number generated in the medium $N_{HH}^{(gen.)}$ we have corrected the $N_{HH}^{(meas.)}$ by estimating the attenuation factor $A$ introduced by absorption effects induced by a single-XUV-photon ionization of Xe i.e. $N_{HH}^{(gen.)}\approx A\times N_{HH}^{(meas.)}$. Considering that the XUV absorption length $L_{abs}=1/(\varrho_{Xe}\sigma^{(1)})$ caused by single--XUV--photon ionization process (where $\sigma^{(1)}\approx 4.3\times10^{-17}$ cm$^2$ is the single photon ionization cross section of Xenon \cite{Samson}) is in the range of $\approx 80$ $\mu$m. Calculating the harmonic photons exiting the medium in case of excluding the absorption effects from the phase matching conditions\cite{Constant} we estimate that $A\sim 10^2 - 10^3$.  Considering that $q$ IR photons are required for the generation of the $q$th harmonic, the number of the IR photons absorbed towards harmonic emission is in the range of $\sim 10^{11}-10^{12}$ photons per pulse (for $q = 15$).

\section{Quantum Spectrometer}

The Quantum Spectrometer (QS) approach relies on shot-to-shot correlation between the photon number of the generated high harmonics (HH) (lying in the plateau and cut-off part of the harmonic spectrum) and the IR field exiting the medium. Detailed description of the approach can be found in refs. \cite{Paris2,Paris3}. Briefly, keeping the energy stability of the driving field at the level of $\approx 1$\%, and after subtracting the electronic noise from each laser shot (as described above), we create the joint XUV--vs--IR photon number distribution using the signal of $i_{HH}$ ($S_{PD_{HH}}$) and $i_{out}$ ($S_{PD_{out}}$) (gray points in Fig. S2a). 
	
\begin{figure*}[h!]
\includegraphics[width=17 cm]{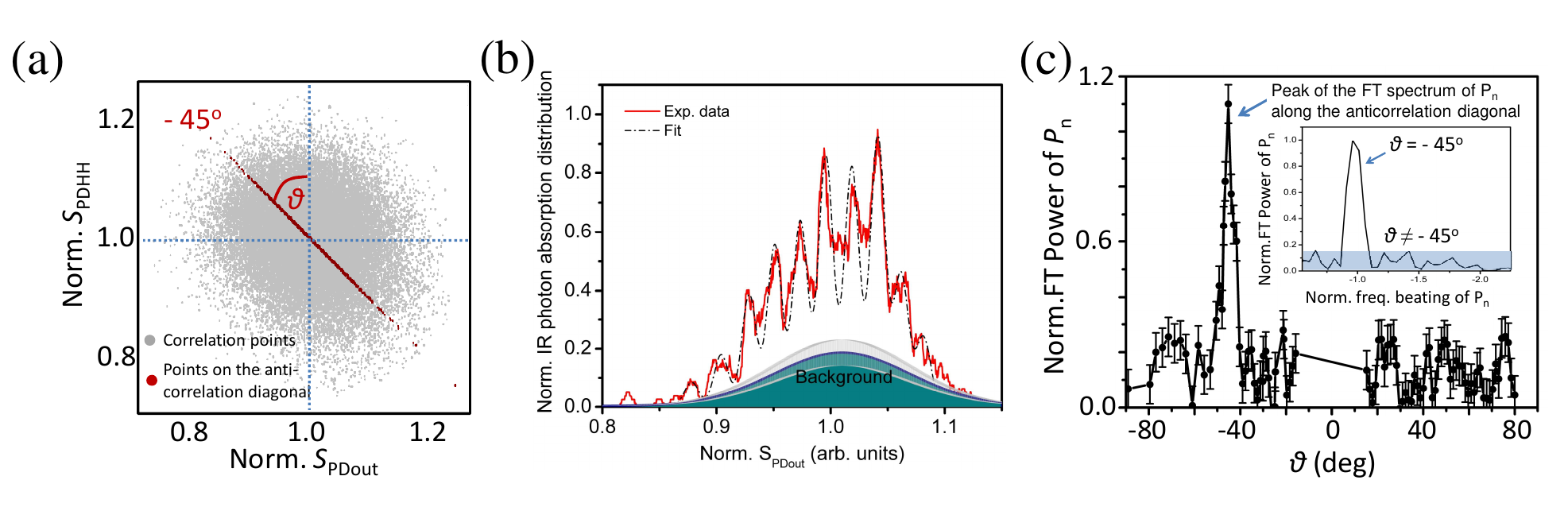}
\caption{\textbf {Quantum spectrometer}. (a) XUV($S_{PD_{HH}}$)--vs--IR($S_{PD_{out}}$) joint photon number distribution (gray points) showing the selected points ($\approx 4$\% of the total number) along the anti--correlation diagonal placed at $\vartheta\approx-45$ deg (red points). (b) Probability to absorb IR photons ($P_n$) towards the harmonic generation (red line). The mean IR photon number detected by the diode was in the range of $\sim 10^{8}$ photons. The black dashed-dot curve is the best fit of an analytical function given by the sum of a sequence of Gaussian functions. The green shaded area shows the background distribution resulted by fitting a Gaussian function on the data obtained by subtracting the minima of the raw data (red line) from the minima of the black dashed-dot curve. The gray bars represent the fit error. (c) Dependence of the Fourier Transform (FT) of $P_n$ on the angle of the anti--correlation diagonal, $\vartheta$. The error bars represent the standard deviation of the mean. The inset shows the peak which corresponds to the periodicity of multi-peak structure of $P_n$ which appears only when $\vartheta\approx-45$ deg, i.e., only along the anti-correlation diagonal. The blue-shaded area in the inset depicts the noise level of the FT.}
\end{figure*}

Both XUV and IR detectors of the QS are operating in the linear regime. Since the intensity dependence of the generated harmonic photons is the same with the IR photon losses, which are associated with all processes (ionization, harmonic generation, ATI etc.) taking place in the interaction region, the mean value of the XUV signal is set to be balanced to the mean value of the IR signal \cite{Paris2,Paris3}. Thus, the joint XUV--vs--IR distribution is round and contains the information of all processes taking place in the interaction region. The points that directly correlate the IR photon losses to the photons generated by the HHG process is a very small portion of the points on the joint XUV--vs--IR distribution and cannot distort the round shape of the joint XUV--vs--IR distribution. After creating the joint XUV--vs--IR photon number distribution, we select only the shots that provide signal along the anti-correlation diagonal placed at $\vartheta\approx-45$ deg of the joint distribution (red points in Fig. S2a). This is the optimum and physically acceptable way (as is based on energy conservation i.e. when the photon number of the XUV increases the photon number of the IR decreases) to disentangle the high-harmonic generation process from all other processes induced by the interaction. These points provide the probability to absorb IR photons ($P_n$) towards the harmonic generation \cite{Paris2, Paris3, Paris1}. In other words, the shot-to-shot photon number distribution of the IR beam exiting the medium (gray points in Fig. S2a) contains the information of all processes induced by the interaction including those that are irrelevant with the harmonic process. By selecting the points along the anti-correlation diagonal we collect only the shots that are relevant to the harmonic emission and we remove the unwanted background associated with all processes irrelevant to the harmonic generation. The width of the anticorrelation diagonal $\delta_{ac}$ defines the level of the conditioning on the HHG process i.e. the smaller the width of the anticorrelation diagonal the better the conditioning on the HHG process. The minimum value of the width is determined by the experimental data using the accuracy to find the peak of the joint distribution which is $\delta_{ac} \equiv w/\sqrt{F}$, (where $w\approx 7$\% is the percentage of the width of the IR photon number relative to the mean value and $F$ is the number of shots). This value also defines the maximum resolution of the QS to resolve the harmonic peaks on the multi-peak structure of IR photon absorption probability distribution. The value $\delta_{ac}$ used in the present work was $\approx 2$\%. This value found to be the optimum in the present experiment, as it provides the optimum visibility and statistical significance of the multi-peak structure of IR photon absorption probability distribution (Fig. S2b). Under these conditions the remaining background distribution associated to processes irrelevant to the HHG process was found to be in the level of $\approx 20$\% of the number of points on the anti--correlation diagonal. This introduces a Gaussian background distribution (green shaded area in Fig. S2b) which needs to be subtracted from the data (see the Quantum Tomography Section 4). The resulted Wigner function is shown in Fig. 4 of the main text of the manuscript. In order to further justify the validity multi-peak structure along the anti-correlation diagonal and exclude the existence of any potential artifacts, we have used a method which relies on the spectrum obtained by Fourier Transform (FT) of the $P_n$ corresponding to different angles $\vartheta$ of the diagonal. The dependence of FT power of $P_n$ on $\vartheta$ is shown in Fig. S2c. A peak, which corresponds to the periodicity of multi-peak structure of $P_n$ (inset of Fig. S2c), appears only when the angle of the diagonal is at $\approx-45$ deg  i.e. only along the anti-correlation diagonal. The width of the peak reflects how quickly the photon distribution loses this modulated form as we move away from the anti-correlation diagonal. We note, that using points of the joint XUV--vs--IR photon number distribution at angles different than the anticorreletion diagonal, is conceptually wrong and against the fundamental principle of energy conservation and the correlation approach.

\section{Quantum Tomography}

The values of of $i_{\varphi}$ used for the characterization of the quantum state are those corresponding to the shots lying along the anti--correlation diagonal. These are directly proportional to the measurement of the electric field operator $\hat{E}_{in} (\varphi) \propto \hat{x}_{\varphi}=\cos(\varphi) \hat{x}+ \sin(\varphi) \hat{p}$. $\hat{x}=(\hat{a}+\hat{a}^\dagger)/\sqrt{2}$ and $\hat{p}=(\hat{a}-\hat{a}^\dagger)/i\sqrt{2}$ are the non-commuting quadrature field operators (namely amplitude and phase quadratures operators), analogues to the position and momentum operators of a particle in an harmonic oscillator and $\hat{a}$, $\hat{a}^\dagger$ are the photon annihilation and creation operators, respectively. Repeated measurements of $\hat{x}_{\varphi}$  at each $\varphi$ (Fig. S3a-c) provides the probability distribution $P_{\varphi}(x_{\varphi})=\langle{x_{\varphi}}|{\hat{\rho}}|{x_{\varphi}}\rangle$ of its eigenvalues $x_{\varphi}$ (where $\hat\rho\equiv|{\phi_{post}}\rangle \langle {\phi_{post}}|$ is the density operator of the light state and $|{x_{\varphi}}\rangle$ the eigenstate of $x_{\varphi}$). For each data set in the range of $0 < \varphi < \pi $ around $\Delta\tau \approx 0$, the Wigner function was reconstructed by means of the inverse Radon transformation implemented via the standard filtered back-projection algorithm\cite{Lvovsky, Breitenbach, Leonhardt}. 
	
\begin{figure*} [h!]
\includegraphics[width= 16 cm]{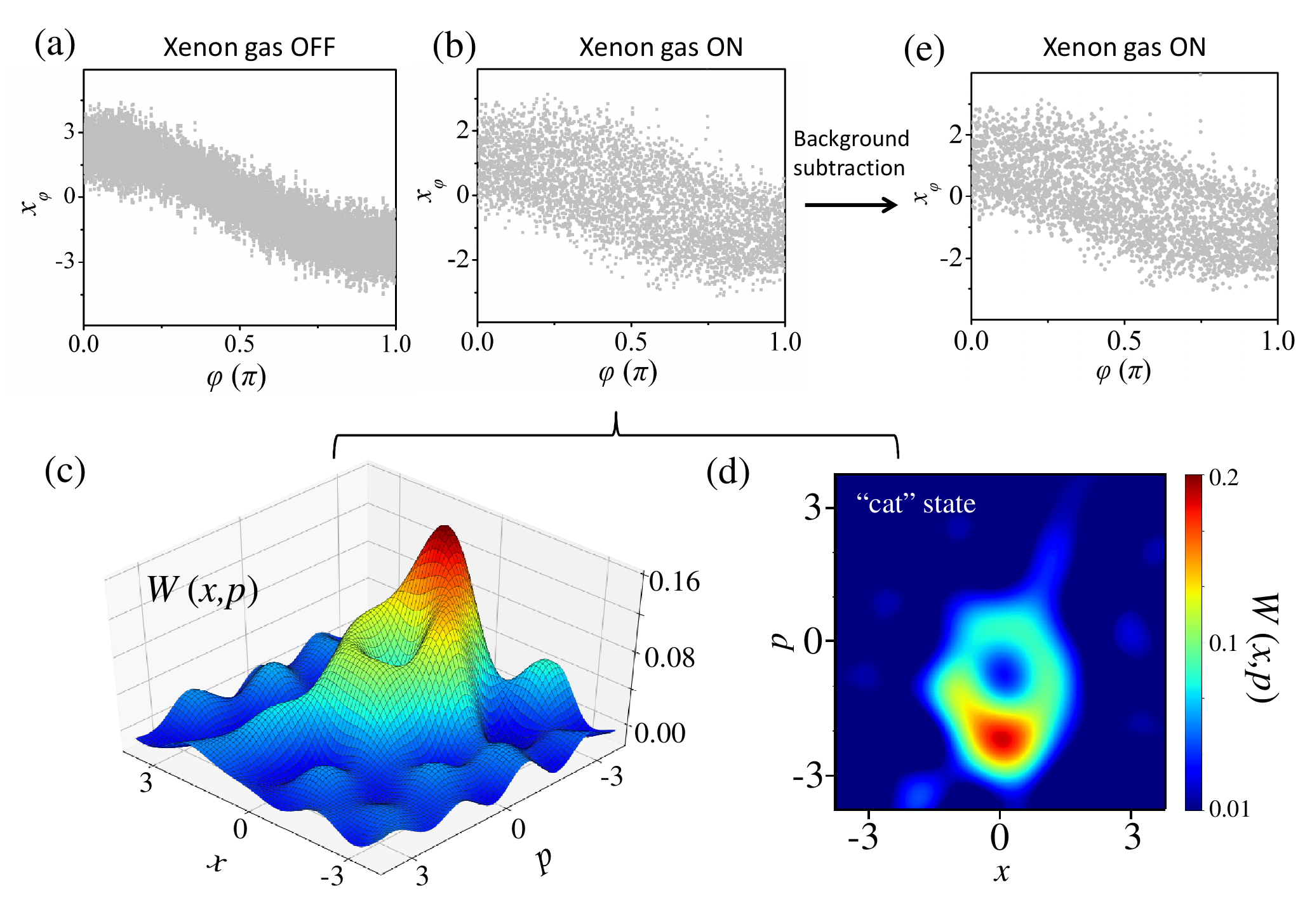}
\caption{\textbf {Field quadratures and reconstruction of the Wigner functions}. (a) and (b) measurement of the field quadratures $\hat{x}_{\varphi}$ as a function of $\varphi$ when the Xenon gas and the QS were switched--off and --on, respectively. The data of (a) have been used for the reconstruction of the Wigner function of the coherent state of the laser field shown in Fig. 3 of the main text of the manuscript. (c),(d) Reconstructed Wigner function $W(x,p)$ using the data of (b). (e) The data of (b) after the subtraction of the background probability distribution shown in Fig. S2b. The data of (e) have been used for the reconstruction of the Wigner function of the ``cat'' state shown in Fig. 4 of the main text of the manuscript.$\it x$, $\it p$ are the values of the quadrature field operators $\hat{x}=(\hat{a}+\hat{a}^\dagger)/\sqrt{2}$ and $\hat{p}=(\hat{a}-\hat{a}^\dagger)/i\sqrt{2}$.}
\end{figure*}
	
The measurement of $\hat{x}_{\varphi}$ shown in Figs. S3a and S3b was performed when the laser-atom interaction was switched--off (i.e. the Xenon gas jet and QS were switched off) and switched--on (i.e. the Xenon gas jet and QS were switched on), respectively. The data of Fig. S3a were used for the reconstruction of the Wigner function of the coherent state of the driving field, shown in Fig. 3b of the main text of the manuscript. The Wigner function reconstructed by the data shown in Fig. S3b is shown in Figs. S3c,d. Although this Wigner function clearly depicts all the features of the optical ``cat'' state, as have been predicted by the theoretical calculations shown in Fig. 2 of the main text of the manuscript, the minimum value at the center of the ring-like distribution is not negative. This is attributed mainly to the remaining background distribution associated to processes irrelevant to the HHG process (green shaded area in Fig. S2b). Subtracting this background from the data shown in Fig. S3b, we obtain the data shown in Fig. S3e, which result to a Wigner function (Fig. 4 of the main text of the manuscript) with negative values at the center of the ring-like distribution as is predicted by the theoretical calculations. Here, we would like to note that, to our knowledge, the photon number of the existing optical ”cat” states is in the range of few photons. These ”cat” states are produced by means of light engineering protocols \cite{Ourjoumtsev, Zavata, Ourjoumtsev1, Wakui} using as primary sources single (or few) photon number light states, vacuum squeezed states and/or detection schemes.
	
In these schemes, the experimental conditions, the primary laser sources, filtering processes as well as the detection systems play an important role towards the optimization of the negativity of the Wigner functions. In the present work the optical "cat" states are not generated using the aforementioned optical engineering approaches. The optical "cat" states are naturally generated by the superposition of the coherent states (which in principle can be of high photon number) which is created by conditioning on HHG the amplitude shifted coherent state. The amplitude attenuation of the coherent state was done in a coherent, non-dissipative manner only to reduce the photon number of the coherent state in order meet the criteria of the QT approach associated with the ratio $E_{r}/E_{in}$. In other words, although in the present work the characterization of the optical ``cat'' states was performed using a mean photon number of $\langle n\rangle \approx 1.98$ photons, the approach is applicable for the generation of optical "cat" states with photon numbers significantly higher than the existing sources. 
	
The algorithm used to reconstruct the Wigner functions was applied directly to the quadrature values $x_{\varphi,k}$ (where $k$ is the index of each value) using the formula\cite{Lvovsky, Leonhardt} $W_{rec}(x,p)\simeq \frac{1}{2\pi^{2}N} \sum_{k=1}^{N}K(x\cdot \cos(\varphi_{k})+p\cdot \sin(\varphi_{k})-x_{\varphi, k})$. $K(z)=\frac{1}{2} \int_{-\infty}^{\infty}|\xi|\exp(i\xi z)\,d\xi$ is called integration kernel with $z=x\cdot \cos(\varphi_{k})+p\cdot \sin(\varphi_{k})-x_{\varphi,k}$. The numerical implementation of the integration kernel requires the replacement of the infinite integration limits with a finite cutoff frequency $k_{c}$. The value of this parameter is chosen so as to reduce the numerical artifacts (rapid oscillations) allowing the details of the Wigner function to be resolved. The value of $k_{c}$ used in the present work was 3.7. An estimation of the error of the reconstructed Wigner function has been obtained by comparing (subtracting) the ideal Wigner function of a coherent state with the Wigner function of a coherent state reconstructed using the experimental parameters (i.e. number of points and $k_{c}$). It is found that the deviation from the ideal case is $\sigma\approx 1$\% resulting to an error of $\pm0.002$. The error of the mean photon number $\langle{n}\rangle$ of the field resulting to the Wigner functions of Figs. 3 and 4 of the main text of the manuscript found to be $\pm0.04$ photons. This was obtained following the aforementioned procedure using the density matrices $\rho_{nm}$ in Fock space $(n,m)$. The mean photon number was obtained by the diagonal elements $\rho_{nn}$ of the $\rho_{nm}$ and the relation $\langle{n}\rangle=\sum n\rho_{nn}$.

\section{On the quantum features of the optical "cat" states}

The measured Wigner function presented in Fig. 4 of the main text of the manuscript and in Fig. S3, depicts a ring-like shape with minimum at the center, features that by definition show the presence of a non-classical light state. This is simply because the measured Wigner deviates from the Wigner function of the coherent light state which has Gaussian form as is shown in Fig. 3 of the main text of the manuscript. Additionally, the Wigner function obtained by the raw data (Fig. S3) depicts the main features of an optical "cat" state. It has a ring-like shape with a minimum at the center. However, the minimum obtained by the raw data was not negative. The reason for this, is the presence of the background noise associated with the limitations of the QS approach (see Quantum Spectrometer Section 3). The background noise subtraction does not influence the non-classical character of the state. It is simply a Gaussian background noise which naturally needs to be subtracted by the data. After doing this, the Wigner function keeps all the main features as were shown in Fig. S3, and reveals its negative values at the minimum. This is shown by comparing the Wigner functions before (shown in Fig. S3) and after noise subtraction (shown in Fig. 4 of the main text of the manuscript).
	
The agreement between the Wigner functions obtained by the theory (Fig. 2 of the main text of the manuscript) and the experiment (Fig. 4 of the main text of the manuscript) is justified in the following way: In both cases (theory and experiment), a) the position of the minimum in phase space found to be at $(x_{min}, p_{min}) \approx (0,-0.75)$; b) the minimum depicts negative values; c) the position of the maximum in phase space found to be at $(x_{max}, p_{max}) \approx (0,-1.8)$; d) the mean diameter of the minimum, defined as $D_{min}=\sqrt{(R_x^2+R_p^2)}$ (where $R_x$ and $R_p$ are the radii of $x$ and $p$ quadratures which have been measured at $[W(x_{max},p_{max})-W(x_{min},p_{min})]/2)$), found to be $D_{min} \approx 1.4$.
	
The deviations between the theoretical calculations and experimental results concern the enhanced negativity and the slightly less pronounce ring-like structure obtained by the theoretical calculations. This is attributed to spatial effects in the HHG region which have not been taken into account in the present theoretical calculations. Although such effects cannot alter the main results captured by the theory, 3-D calculations considering propagation effects in the medium using multimode driving field can, and probably will, provide more accurate theoretical results.
	
Finally, we note that in order to obtain the Wigner functions shown in Fig. 2, $\zeta = \langle \alpha| \alpha + \delta \alpha \rangle$ needs to be different from zero, i.e., to have a non-vanishing overlap. Therefore, if these two states ($|\alpha\rangle$ and $|\alpha + \delta \alpha\rangle$) are fairly similar, as it happens to be, we would not be able to distinguish a peak belonging to $|\alpha\rangle$ and another one to $|\alpha + \delta \alpha\rangle$. For that to happen, we need both states to be "distinguishable", a limit that we cannot reach because the state after the QS/QT approach is $|\alpha + \delta \alpha\rangle - \zeta |\alpha\rangle$, and in the limit where both states are "distinguishable" we simply get a coherent state. However, the Wigner function has the natural capability of showing quantum interference effects in terms of negative regions, and as $\delta \alpha$ is different from zero, we are able to observe them.


\begin{thebibliography}{50}
\bibitem{Mourou}
Mourou, G. Nobel Lecture: Extreme light physics and application. {\it Rev. Mod. Phys.} {\bf 91}, 030501 (2019).
		
\bibitem{Strickland}
Strickland, D. Nobel Lecture: Generating high-intensity ultrashort optical pulses. {\it Rev. Mod. Phys.} {\bf 91}, 030502 (2019).
		
\bibitem{Glauber2}
Glauber, R. J. Nobel Lecture: One hundred years of light quanta. {\it Rev. Mod. Phys.} {\bf 78}, 1267–1278 (2006).
		
\bibitem{McPherson1987}
McPherson, A. et al. Studies of multiphoton production of vacuum-ultraviolet radiation in the rare gases. {\it J. Opt. Soc. Am. B} {\bf 4}, 595601 (1987).
		
\bibitem{Ferray1988}
Ferray, M. et al. Multiple-harmonic conversion of 1064 nm radiation in rare gases. {\it J. Phys. B: At. Mol. Opt. Phys.} {\bf 21}, L31 (1988).
		
\bibitem{Maiman}
Maiman, T. H. Stimulated Optical Radiation in Ruby. {\it Nature} {\bf 187}, 493, (1960).
		
\bibitem{Symphony}
Amini, K. et al. Symphony on strong field approximation. {\it Rep. Prog. Phys.}, {\bf 82}, 116001 (2019).
		
\bibitem{Delone}
N.B. Delone and V.P. Krainov, Multiphoton Processes in Atoms (2nd Ed.) (Springer Series on Atomic, Optical, and Plasma Physics, 200).
		
\bibitem{Protopapas}
M. Protopapas, C.H. Keitel and P.L. Knight, Atomic physics with super-high intense lasers {\it Rep. Prog. Phys.}, {\bf 60}, 389-486 (1997).
		
\bibitem{Corkum1993}
Corkum, P. B. Plasma perspective on strong field multiphoton ionization. {\it Phys. Rev. Lett.} {\bf 71}, 1994-1997 (1993).
		
\bibitem{Kulander1993}
Kulander, K. C., Schafer, K. J., and Krause, J. L. Dynamics of short-pulse excitation, ionization and harmonic conversion. (Super-Intense Laser Atom Physics, volume 316 of NATO Advanced Studies Institute Series B: Physics, pages 95-110, Plenum, New York, 1993).
		
\bibitem{Lewenstein1994}
Lewenstein, M. et al. Theory of high-harmonic generation by low-frequency laser fields. {\it Phys. Rev. A} {\bf 49}, 2117-2132 (1994).
		
\bibitem{Knight}
Knight, P. L., and Miller, A.  Measuring the quantum state of light. (Cambridge Univ. Press, 1997).
		
\bibitem{Ourjoumtsev}
Ourjoumtsev, A., Jeong, H., Tualle--Brouri, R. and Grangier, P. Generation of optical `Schr\"{o}dinger cats' from photon number states. {\it Nature} {\bf 448}, 784 (2007).
		
\bibitem{Zavata}
Zavatta, A., Viciani, S., Bellini, M., Quantum--to--classical transition with single--photon--added coherent states of light. {\it Science} {\bf 306}, 660 (2004).
		
\bibitem{Ourjoumtsev1}
Ourjoumtsev, A., Tualle-Brouri, R., Laurat, J. and Grangier, P. Generating optical Schr\"{o}dinger kittens for quantum information processing. {\it Science} {\bf 312}, 83–86 (2006).
		
\bibitem{Wakui}
Wakui, K., Takahashi, H., Furusawa, A. and Sasaki, M. Controllable generation of highly nonclassical states from nearly pure squeezed vacua. {\it Opt. Express} {\bf 15}, 3568–3574 (2007).
		
\bibitem{Acin}
Ac\'{\i}n, A. et al. The quantum technologies roadmap: a European community view. {\it New J. Phys.}, {\bf 20} 080201 (2018).
		
\bibitem{Walmsay}
Walmsley, I. A. Quantum optics: Science and technology in a new light. {\it Science}, {\bf 348}, 525 (2015).
		
\bibitem{Deutsch}
Deutsch, I. H. Harnessing the Power of the Second Quantum Revolution. {\it PRX Quantum}, {\bf 1}, 020101 (2020).
		
\bibitem{Diestler2008}
Diestler, D. J. Harmonic generation: Quantum-electrodynamical theory of the harmonic photon-number spectrum. {\it Phys. Rev. A} {\bf 78}, 033814 (2008).
		
\bibitem{Paris1}
Gonoskov, I. A., Tsatrafyllis, N., Kominis, I. K., and Tzallas, P. Quantum optical signatures in strong-field laser physics: Infrared photon counting in high-orderharmonic generation {\it Sci. Rep.} {\bf 6}, 32821 (2016).
		
\bibitem{Foldi}
Gombk\"{o}t\H{o}, \'{A}., Varr\'{o}, S., Mati, P., and F\"{o}ldi, P. High-order harmonic generation as induced by a quantized field: Phase-space picture. {\it Phys. Rev. A} {\bf 101}, 013418 (2020).
		
\bibitem{Gorlach2019}
Gorlach, A., Neufeld, O., Rivera, N., Cohen, O., and Kaminer, I. The quantum-optical nature of high harmonic generation. {\it Nat. Commun} {\bf 11}, 4598 (2020).
		
\bibitem{Yangaliev}
Yangaliev, D. N., Krainov, V. P., and Tolstikhin, O. I. Quantum theory of radiation by nonstationary systems with application to high-order harmonic generation. {\it Phys. Rev. A} {\bf 101}, 013410 (2020).
		
\bibitem{Paris2}
Tsatrafyllis, N., Kominis, I. K., Gonoskov, I. A., and Tzallas, P. High-order harmonics measured by the photon statistics of the infrared driving-field exiting the atomic medium. {\it Nat. Commun.} {\bf 8}, 15170 (2017).

\bibitem{Paris3}
Tsatrafyllis, N. et al. Quantum Optical Signatures in a Strong Laser Pulse after Interaction with Semiconductors. {\it Phys. Rev. Lett.}, {\bf 122}, 193602 (2019).

\bibitem{Lvovsky}
Lvovsky, A. I. and Raymer, M. G., Continuous--variable optical quantum--state tomography {\it Rev. Mod. Phys.} {\bf 81}, 299 (2009).
		
\bibitem{Schleich}
Schleich, W. P. Quantum Optics in Phase Space. (Wiley-VHC Verlag, Berlin, 2001).
		
\bibitem{Breitenbach}
Breitenbach, G., Schiller, S. and Mlynek, J. Measurement of the quantum states of squeezed light. {\it Nature} {\bf 387}, 471 (1997).
		
\bibitem{Krausz2009}
Krausz, F., and Ivanov, M. Y. Attosecond physics. {\it Rev. Mod. Phys.} {\bf 81}, 163-234 (2009).
		
\bibitem{Nayak}	
Nayak, A. et al. Saddle point approaches in strong field physics and generation of attosecond pulses. {\it Phys. Rep.} {\bf 833}, 1-52 (2019).





\bibitem{Wuensche}
W\"{u}nsche, A. {\it J. Opt. B: Quantum Semiclass. Opt.}, {\bf 6}, S47-S59 (2004).

\bibitem{Grynberg}
Grynberg, G., Aspect, A. and Fabre, C. {\it Introduction to Quantum Optics} (Cambridge University Press, Cambridge, 2010).




\bibitem{Altucci}
Altucci, C. {\it et al.}, Characterization of pulsed gas sources for intense laser field--atom interaction experiments {\it J. Phys. D: Appl. Phys.} {\bf 29}, 68 (1996).

\bibitem{Leonhardt}
Leonhardt, U., Knight, P. and Miller, A. {\it Measuring the Quantum State of Light} (Cambridge University Press, 1997).
		
\bibitem{Samson}
Samson, J. A. R., and Stolte, W. C. Precision measurements of the total photoionization cross-sections of He, Ne, Ar, and Xe. {\it J. Electron. Spec. Rel. Phenom.} {\bf 123}, 265 (2002).
		
\bibitem{Constant}
Constant, E. {\it et al.}, Optimizing High Harmonic Generation in Absorbing Gases: Model and Experiment. {\it Phys. Rev. Lett.} {\bf 82}, 1668 (1999).
		
\end{thebibliography}
\end{document}